\documentstyle[epsfig,wrapfig]{aipproc}

\newcommand{\be}{\begin{equation}}
\newcommand{\ee}{\end{equation}}
\newcommand{\ba}{\begin{eqnarray}}
\newcommand{\ea}{\end{eqnarray}}

\newcommand{\bi}[1]{\bibitem{#1}}
\newcommand{\lab}[1]{\label{#1}}
\begin{document}
\title{
Properties of the Spin-flip Amplitude of Hadron Elastic
Scattering \\ and Possible Polarization Effects at RHIC
 }
\author{O.~V.~Selyugin\footnote{E-mail: selugin@thsun1.jinr.ru}
}
\address{
Joint Institute for Nuclear Research, Dubna, 141980 Russia}

\maketitle

\begin{abstract}
With relation to  the RHIC spin program we research the polarization
effects in
elastic proton-proton scattering at small momentum transfer and in the
diffraction dip region. The calculations take into account the Coulomb-hadron
interference effects including the additional Coulomb-hadron phase. In
particular we show the impact of the form of the hadron potential  at large
distances on the behavior of the hadron spin-flip amplitude at small angles.
The $t$-dependence of the spin-flip amplitude of  high energy hadron elastic
scattering is analyzed under different assumptions on the hadron interaction.
\end{abstract}

\section{Introduction}
 Several attempts to extract the spin-flip amplitude from the experimental
data  show that the ratio of spin-flip to spin-nonflip
amplitudes can be non-negligible and may be only slightly dependent on energy
\cite{akch,sel-pl}.

 For the definition of new effects at small angles
  and especially in the region of the diffraction minimum
  one must  know the effects of the Coulomb-hadron interference
 with sufficiently high accuracy.
  The Coulomb-hadron phase was calculated
 in the entire diffraction domain taking into account  the form factors
 of the  nucleons \cite{prd-sum}.
  Some polarization effects connected with the Coulomb hadron
   interference, including some possible  odderon contribution,
   were also calculated \cite{z00}.

   The model-dependent analysis based on all the existing experimental
   data of the spin-correlation parameters above $p_L \ \geq
 \  6 \ GeV$
   allows  us to determine the structure of the hadron spin-flip
   amplitude at high energies and to predict its behavior at
   superhigh energies \cite{yaf-wak}. This analysis shows
   that the ratios
   $Re \ \phi^{h}_{5}(s,t) / (\sqrt{|t|} \ Re \ \phi^{h}_{1}(s,t))$ and
   $Im \ \phi^{h}_{5}(s,t)/(\sqrt{|t|} \ Im \ \phi^{h}_{1}(s,t))$
   depend on $\ s$ and $t$ (see Fig.1 a,b). At small momentum transfers,
   it was found that the slope of the ``residual'' spin-flip
    amplitudes is approximately
   twice the slope of the spin-non flip amplitude.
       The obtained spin-flip amplitude leads to the additional contribution
  to the pure CNI effect at small $t$ (Fig. 1 c).

\begin{figure}[t!]
\vspace{-8mm}
\begin{center}
\begin{tabular}{ccc}
\mbox{\epsfig{figure=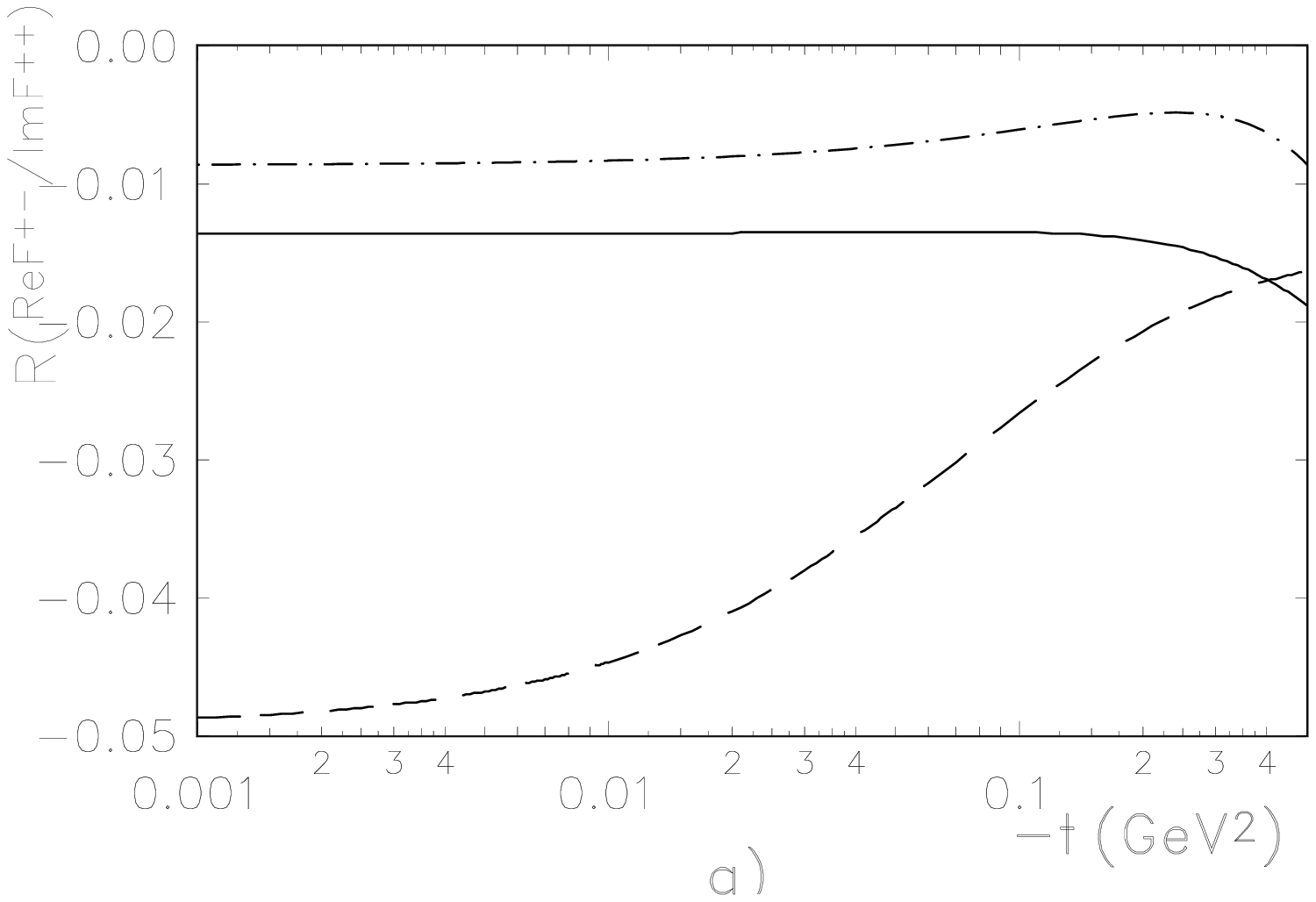,width=4.5cm,height=4cm}}&
\mbox{\epsfig{figure=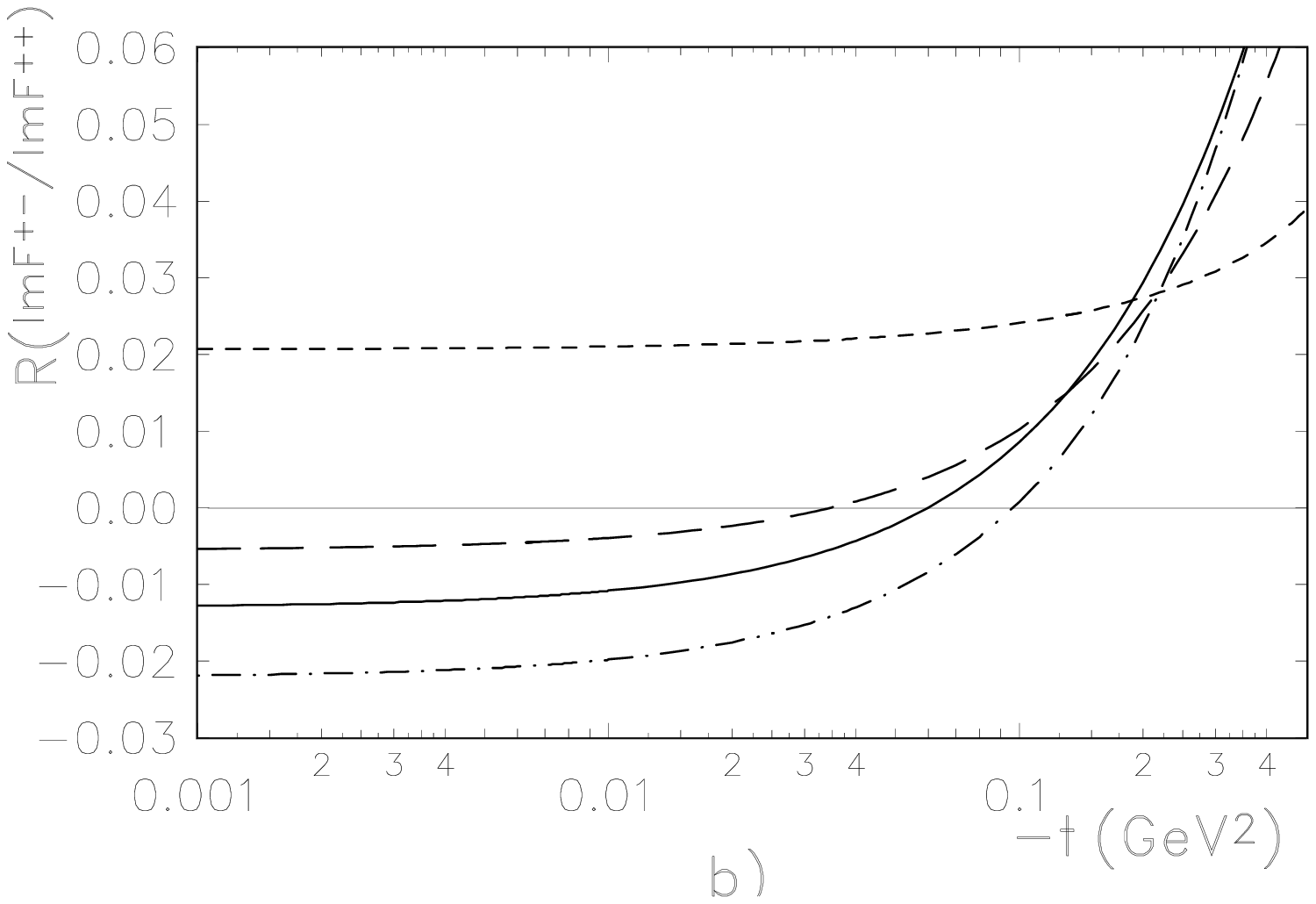,width=4.5cm,height=4cm}}&
\mbox{\epsfig{figure=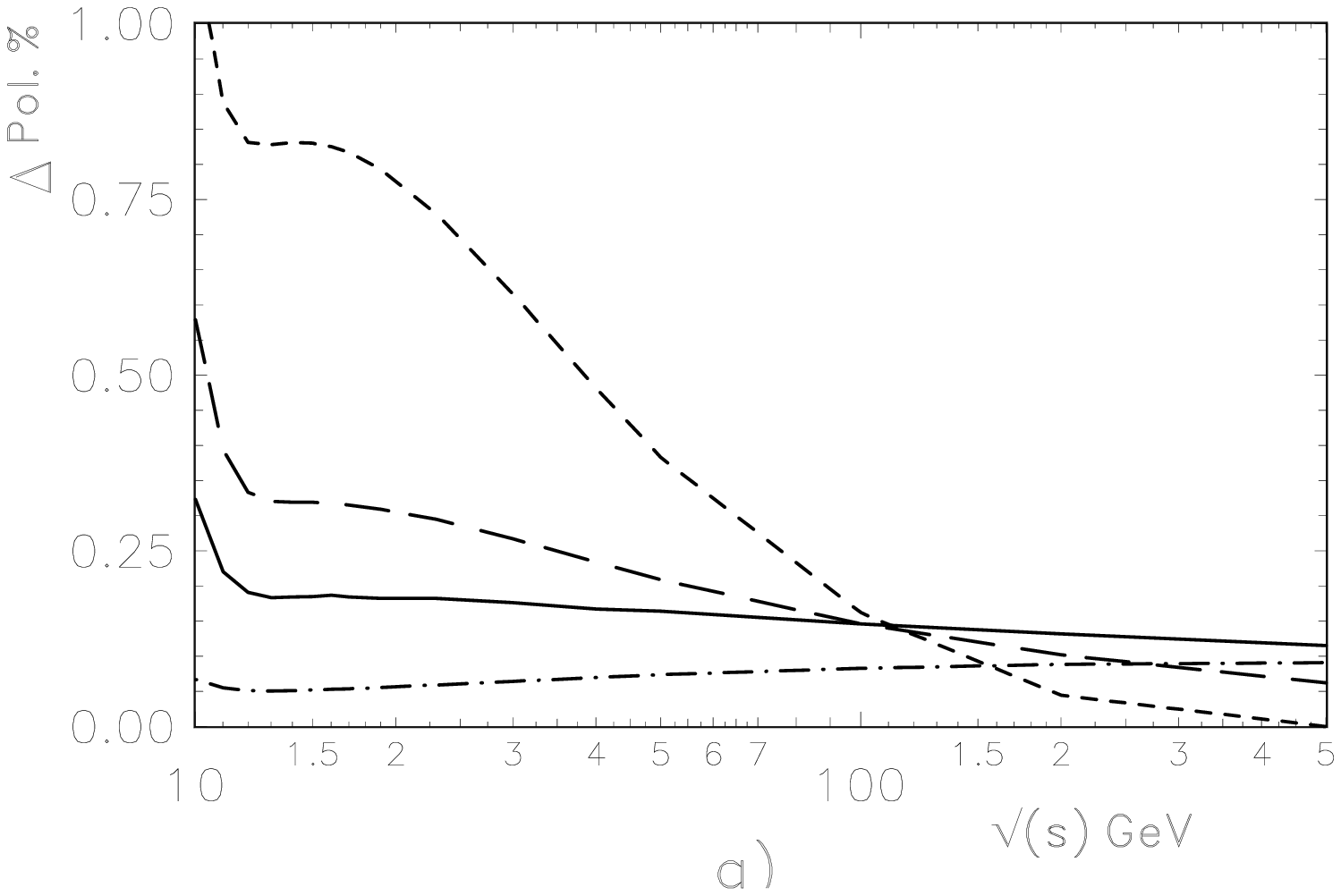,width=4.5cm,height=4cm}} \\
{\bf (a)} & {\bf (b)} &  {\bf (c)}
\end{tabular}
\caption{
 ({\bf a} ) and {\bf (b))}  Ratio of the imaginary (a) and real (b) part
      of the  "residual" $F_{h}^{+-}$
  $. \ \ $ \hspace{2.cm} to the imaginary $F_{h}^{++}$ }
({\bf    c})  {\small    Contribution to the pure CNI effect from
        the model $F_{h}^{+-}$}
     (the our calculations  at $t_{max}$, $|t| = 0.001 \ GeV^2$
      $|t| = 0.01 \ GeV^2$,$|t| = 0.1 \ GeV^2$
       are shown
         the full line,
         the long dashed line,
         the  dashed line and
         the  doted line respectively.
\end{center}
\end{figure}

   The dependence of the hadron spin-flip amplitude on $t$
  at small angles is closely related with the basic structure of
  hadrons at large distances. We show that
  the slope of the so-called ``reduced'' hadron spin-flip amplitude
  (the hadron spin-flip amplitude without the kinematic factor $\sqrt{|t|}$)
  can be larger
  than the slope of the hadron spin-non-flip amplitude,
  as was observed long ago \cite{predaz}.

 \section{The slope of the hadron  amplitudes}

     For an exponential form of the amplitudes this  coincides
    with the usual slope of the differential cross sections divided by $2$.
 At small $t$ ($\sim 0 \div 0.1 \ GeV^2$), practically all
 semiphenomenological analyses assume:
 $ B_{1}^{+} \ \approx \ B_{2}^{+} \ \approx \
    B_{1}^{-} \ \approx \ B_{2}^{-}  $.
  If the potentials $V_{++}$ and $V_{+-}$ are
   assumed to have a Gaussian form
  in the first Born approximation
  $\phi^{h}_{1}$ and $\hat{\phi_h}^{5}$
   will have  the same form
$\phi^{h}_{1}(s,t) \sim  exp({-B \Delta^{2}})$,
$\phi^{h}_{5}(s,t)
 \sim
= \ q \ \ B \ exp({-B \Delta^{2}})$.
  In this special case, therefore, the slopes of
 the  spin-flip and  ``residual''spin-non-flip amplitudes are
  indeed the same.
 A Gaussian form of the potential is adequate to represent
 the central part of the  hadronic  interaction.
  The form cuts off the Bessel function
  and the contributions at large distances.
  If, however, the potential (or the corresponding eikonal)
 has a long tail (exponential or power)
  in the impact parameter,  the Bessel functions
  can not be taken in the  approximation form
 and  the full integration leads to  different results.

 If we take
 $ \chi_{i}(b,s) \ \sim \ H \ e^{- a \ \rho}, $
we obtain
\ba
 F_{nf} (s,t) =  a/[(a^2 + q^2)^{3/2}] \approx \
                  1/[a \sqrt{a^2+q^2}] \ exp({-B q^2})
\ea
with $B \ = \ 1/a^{2}$.
 For the ``residual'' spin-flip amplitude,
  on the other hand, we obtain  \cite{tur1}
\ba
\sqrt{|t|} \tilde{F_{sf}}(s,t) =  (3 \ a \ q)/[(a^2 + q^2)^{5/2}] \
  \approx \ (3 \ a  q \ B^{2})/( \sqrt{a^2+q^2}) \ \ exp({-2 \ B q^2}).
\ea
  In this case, therefore,
  the slope of the ``residual'' spin-flip amplitude exceeds the slope
 of the spin-non-flip amplitudes by a factor of two.
  A similar behaviour can be obtained with the standard dipole form factor
  \cite{tur1}.

\section{The determination of the structure of the hadron spin-flip amplitude}

  Note that if the ``reduced'' spin-flip amplitude
  is not small,  the impact of a large $B^{-}$ will reflect
   in the behavior of the differential cross section at small
   angles \cite{sel-sl}.
   The method gives  only  the absolute value of the coefficient
  of the  spin-flip amplitude. The imaginary
  and real parts of the spin-flip amplitude can be found only from
  the measurements of the spin correlation coefficient.

 Let us
take the spin nonflip
 amplitude in the standard exponential form with
  definite parameters:
 slope $B^{+}$, $\sigma_{tot}$ and $\rho^{+}$. For the ``residual''
 spin-flip amplitude, on the other hand,
  we consider  two possibilities:  equal slopes
 $B^{-}=B^{+}$ and  $B^{-}= 2 B^{+}$.
 The results of these  two different
 calculations are shown in Fig.2. It is clear that around
 the maximum of the Coulomb-hadron interference, the difference
 between the two
 variants is very small. But when $|t| \ > 0.01 \ GeV^2$,
  this difference grows. So, if we  try to find the
  contribution of the pomeron spin-flip, we should take into account
  this effect.
    As the value of $A_N$ depends on the determination of the beam
 polarization, let us calculate  the derivative of  
 $A_N$ with respect to $t$,
 for example, at $\sqrt{s} = 500 \ GeV$.

   If we know the parameters of the hadron spin non-flip amplitude,
 the measurement of the analyzing power at small transfer momenta
 helps us to find the structure of the hadron spin-flip amplitude.
  There is a specific point of  the differential cross sections
  and of $A_N$ on the axis of the momentum transfer, - $t_{re}$,
 where
 $|Re F_{c}^{++}| = |Re F_{h}^{++}|$.
  This point $t_{re}$  can be found from
  the  measurement of
 the   differential cross sections   \cite{addapr}.
 At high energies at the point  $t_{re}$ \cite{tur1}
 we obtain for $pp$-scattering
 \ba
   Re F_{sf}^{h}(s,t) = \frac{-1}{2  (Im F_{nf}^{h}(s,t)+Im F_{nf}^{c}(t))}
   A_{N}(s,t) \
     \frac{d\sigma}{dt} \
 	       - Re F_{sf}^{c}(t).   \lab{refhm}
 \ea
 We can again take the hadron spin-nonflip
 and spin-flip
  amplitudes with  definite parameters and calculate the magnitude of
 $A_N$ by the usual complete form
 while  the real part of the hadron
 spin-flip amplitude is given by (\ref{refhm}).
 Our calculation by this formula and the input real part of the
  spin-flip amplitude are shown in Fig. 2 c.
  At the point $t_{re}$ both curves coincide. So if we obtain from
 the accurate measurement  of the differential
 cross sections the value of $t_{re}$, we can find from
   $A_N$ the value of the real part of the hadron spin-flip
 amplitude at the same point of momentum transfer.

\begin{figure}[t!]
\vspace{-8mm}
\begin{center}
\begin{tabular}{ccc}
\mbox{\epsfig{figure=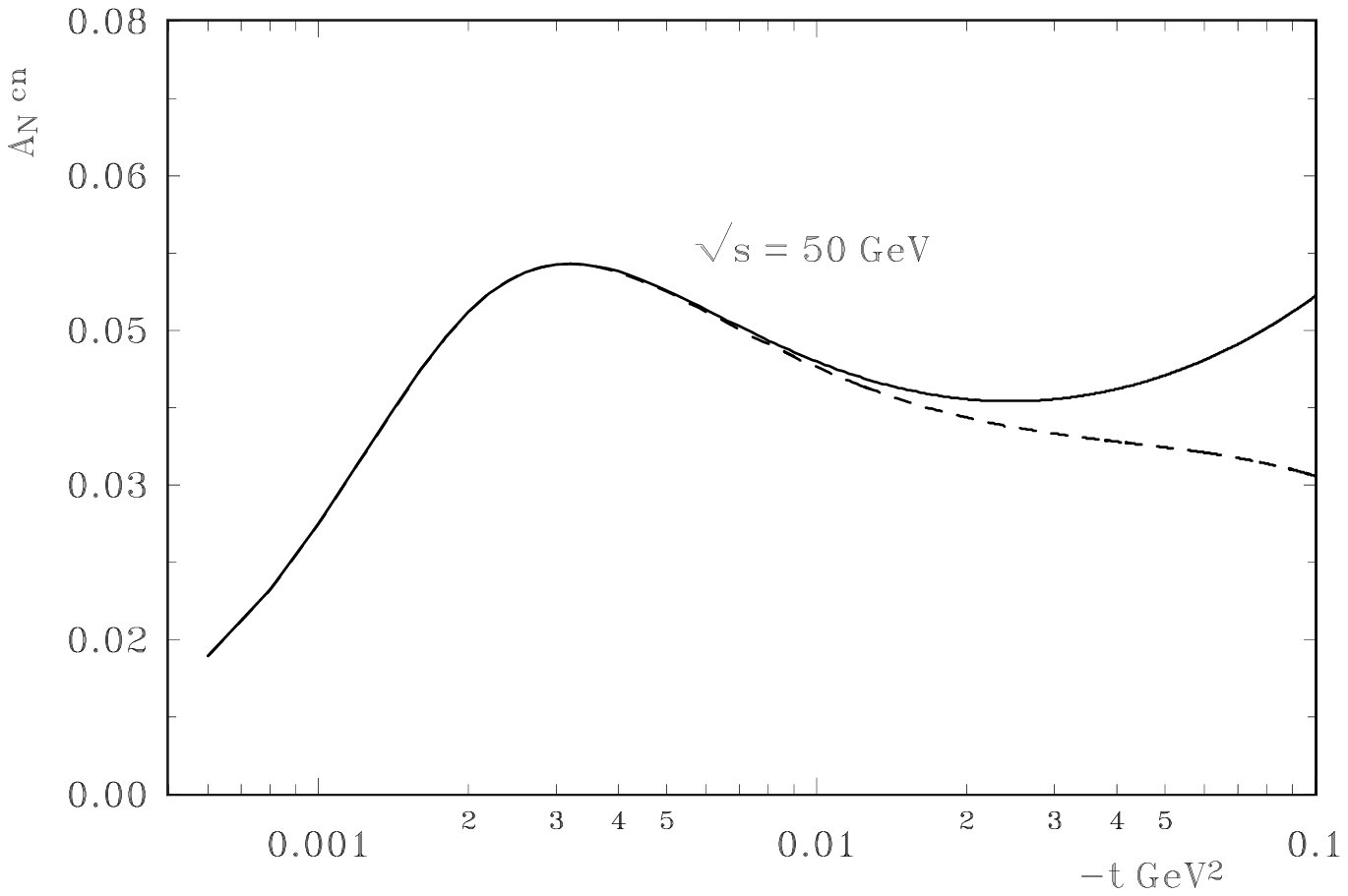,width=4.5cm,height=4cm}}&
\mbox{\epsfig{figure=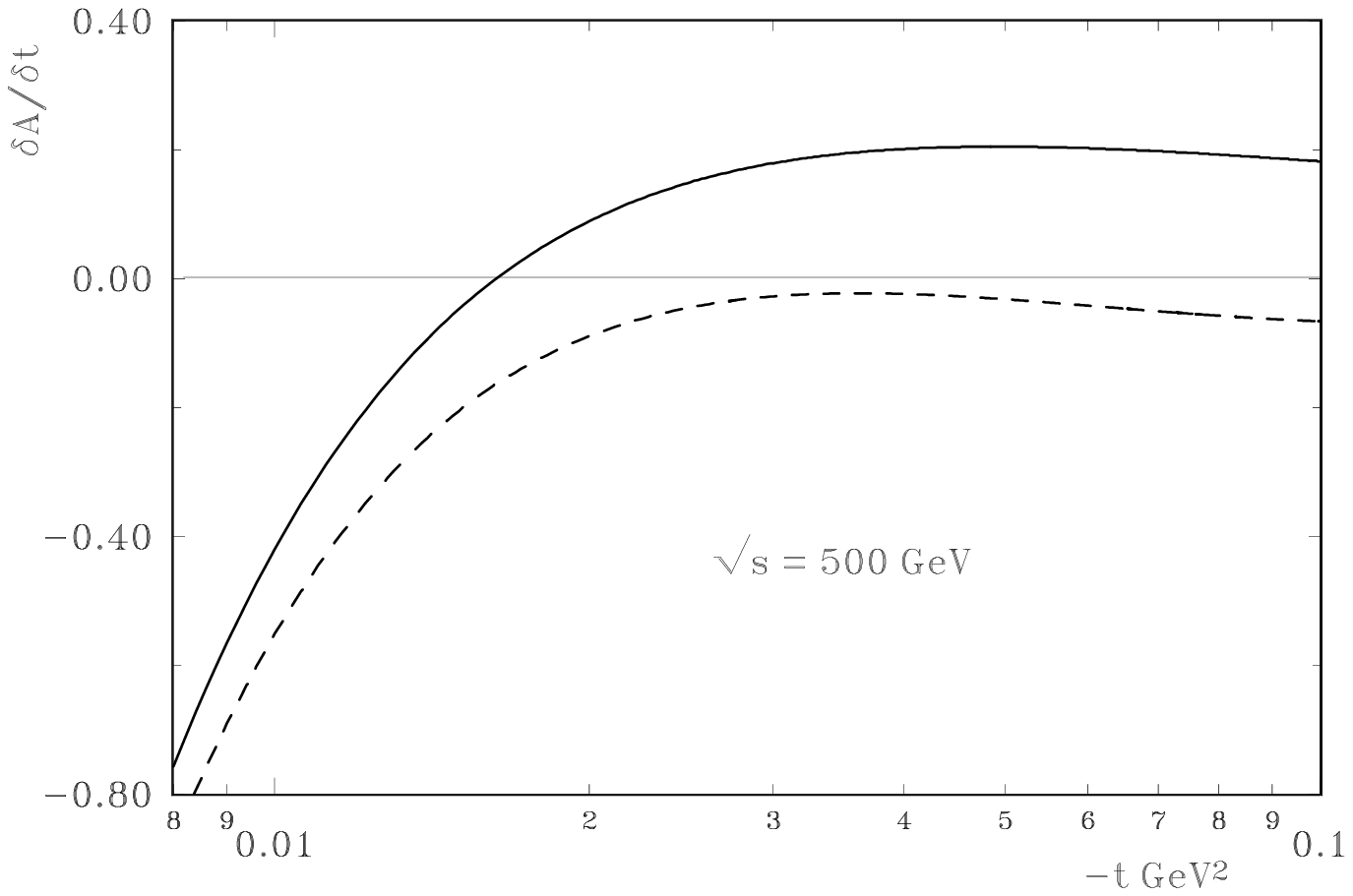,width=4.5cm,height=4cm}}&
\mbox{\epsfig{figure=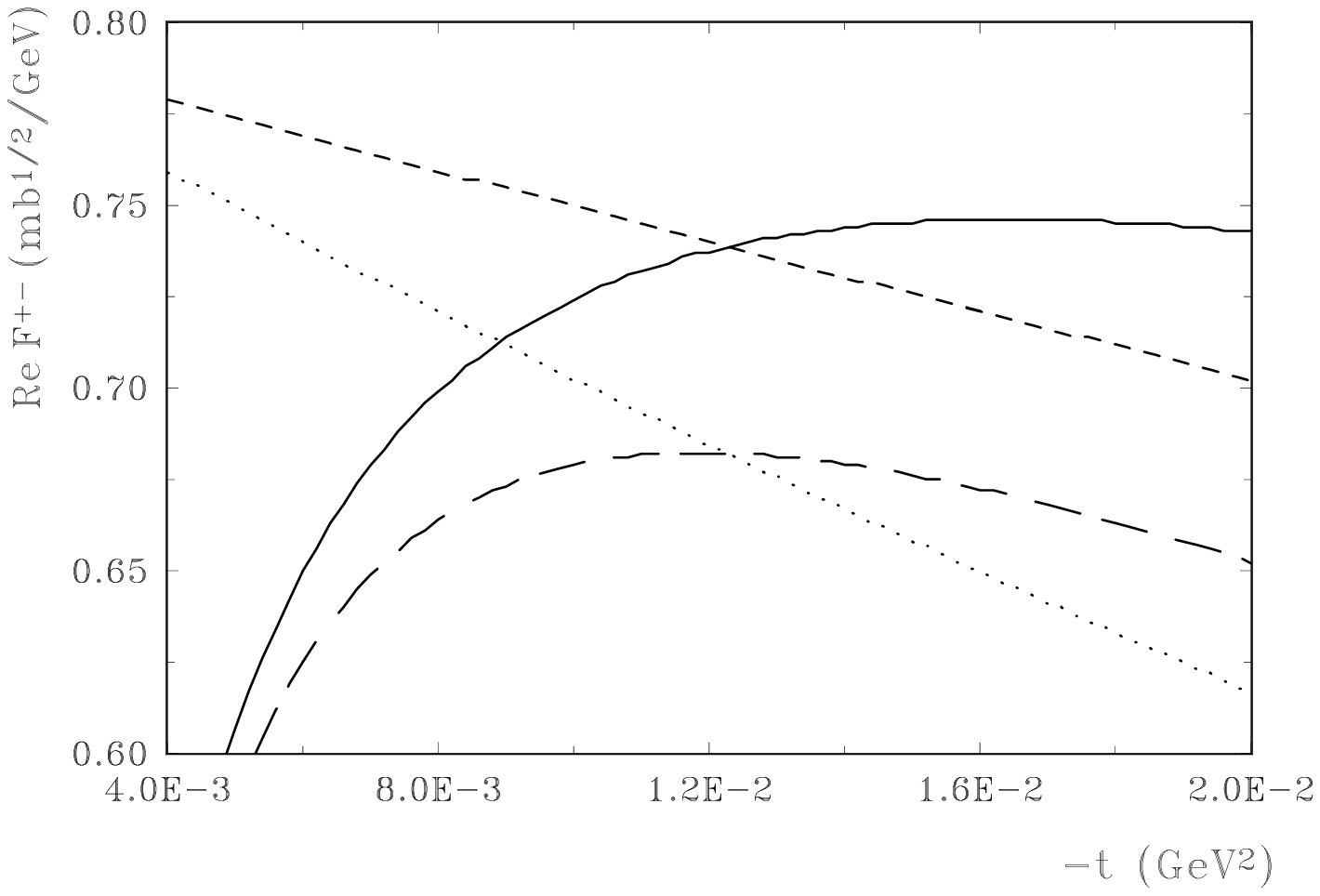,width=4.5cm,height=4cm}} \\
{\bf (a)} & {\bf (b)} &  {\bf (c)}
\end{tabular}
\caption{
({\bf a})
 $A_N $ at $\sqrt{s} = 50 \ GeV$
 ({\bf b})
$- \delta A_N/\delta t$
  at $\sqrt{s} = 500 \ GeV$
  (the solid line is
  with the slope $B_{1}^{-}$  of $\tilde{F_{sf}} $ equal to the slope
   $B_{1}^{+}$ of $F_{nf}$;
   the dashed line is
  with the $B_{1}^{-} \ = \ 2 \ B_{1}^{+} $ .
    ({\bf c})
 The form of $Re ( F^{sf})$ :
 solid and long-dashed lines are calculations by (3);
  short-dashed  and
  dottes lines are model amplitudes
 with $B_{1}^{-} = B_{1}^{+}$ and
  $B_{1}^{-} = 2 \ B_{1}^{+}$, respectively. }
\end{center}
\end{figure}

\section{The model predictions}

\begin{figure}[b!]
\vspace{-8mm}
\begin{center}
\begin{tabular}{ccc}
\epsfig{figure=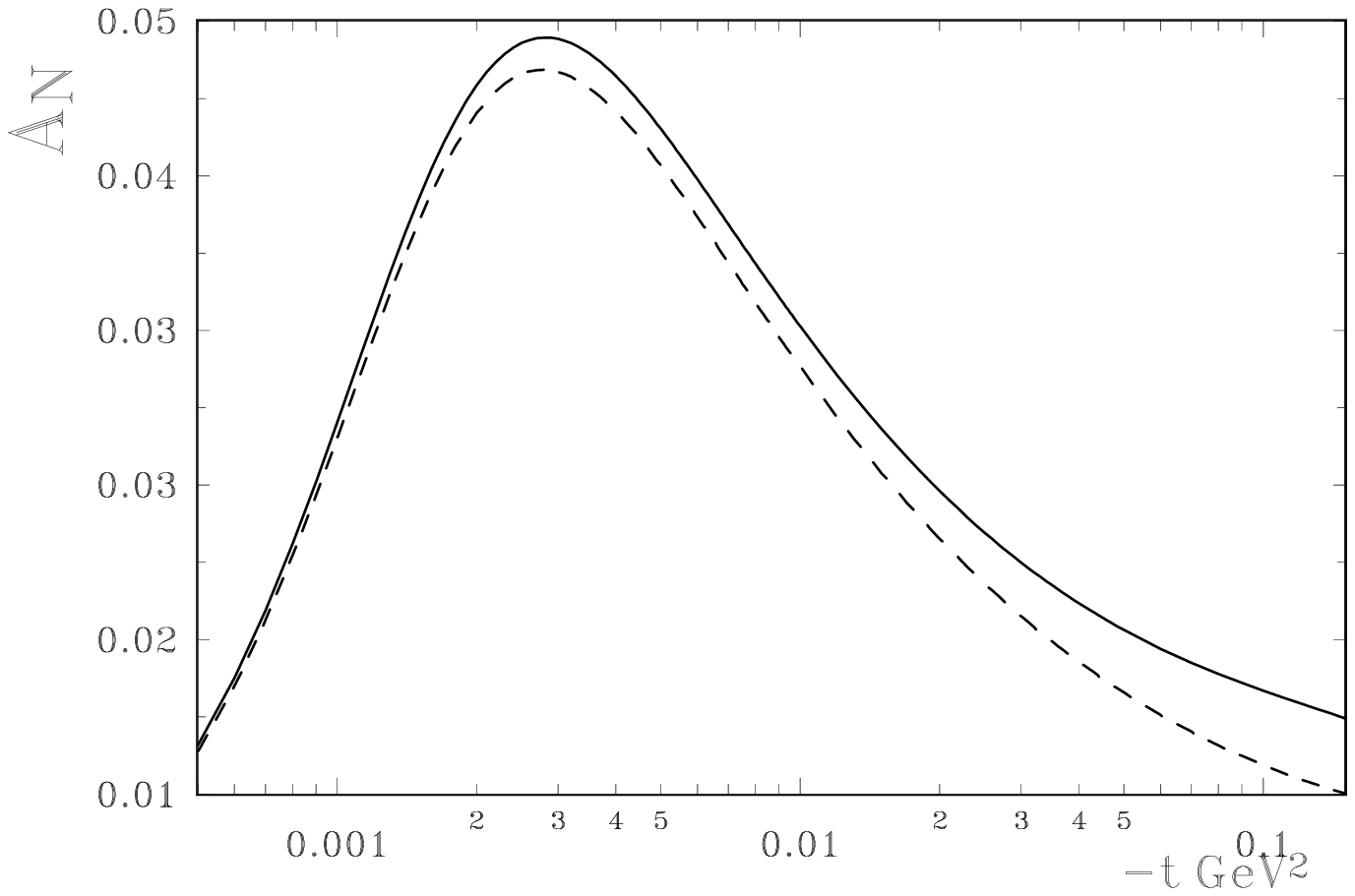,width=4.5cm,height=4cm}&
\epsfig{figure=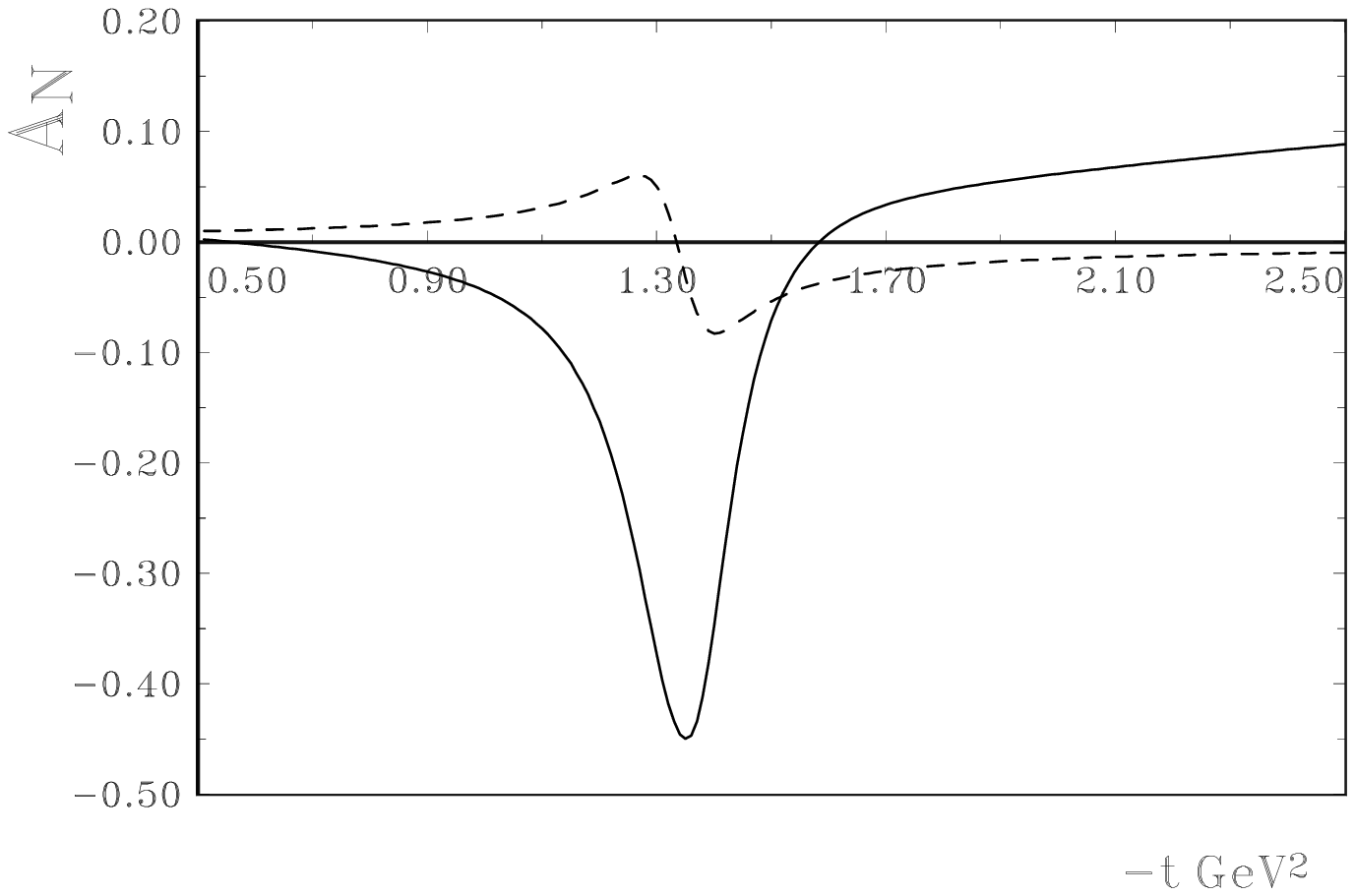,width=4.5cm,height=4cm}&
\epsfig{figure=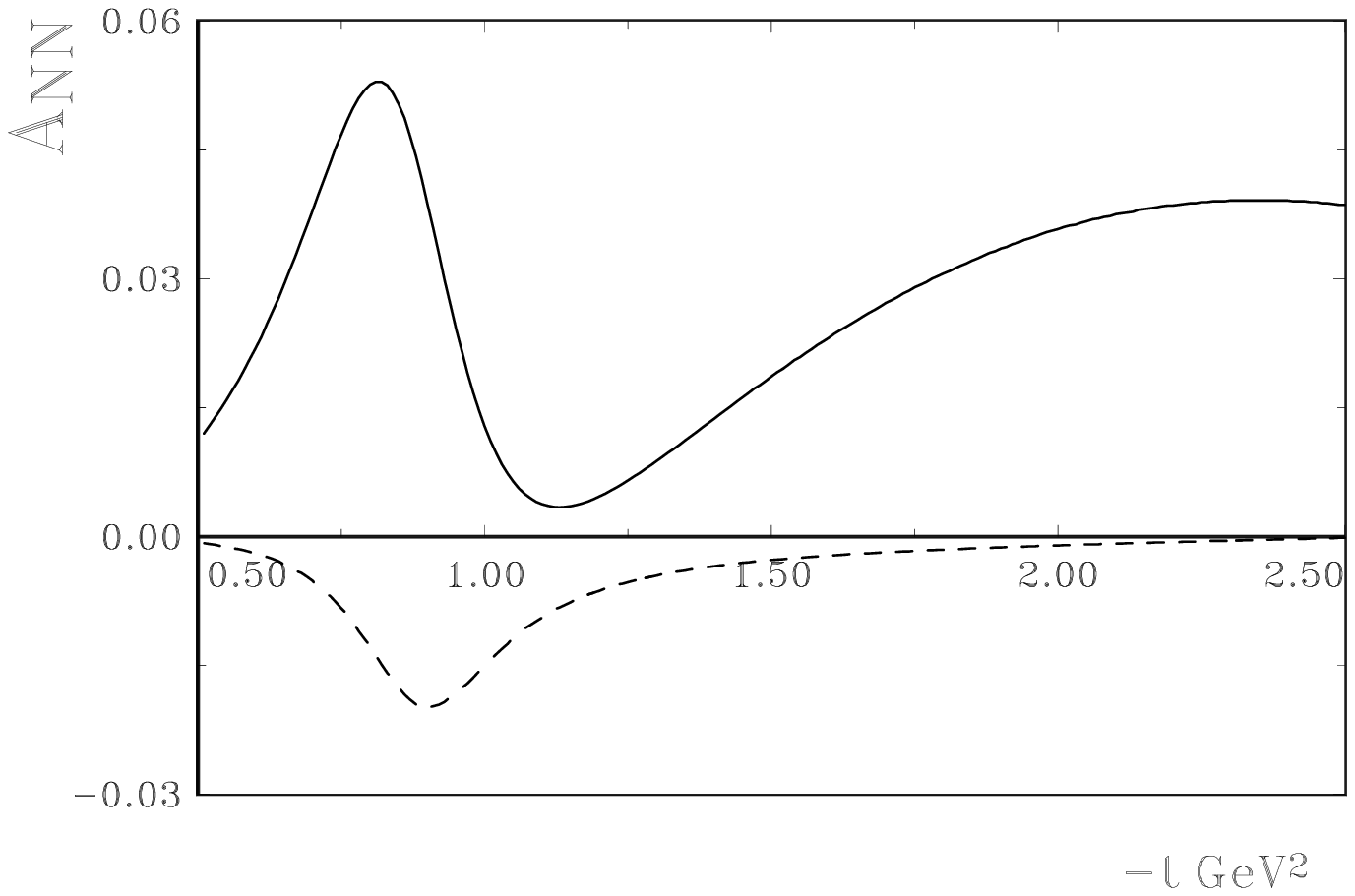,width=4.5cm,height=4cm}\\
{\bf(a)}& {\bf(b)} & {\bf(c)}
\end{tabular}
\caption{
({\bf a}) and ({\bf b})
     $A_N$
 at $\sqrt{s} = 50 \ GeV$ (a)
    the full line is the total $A_N$; the dushed line is the $A_{N}^{CN}$.
 {\bf(c).}
     $A_{NN}$
 at $\sqrt{s} = 500 \ GeV$
	  in the domail of the dip  ;
	  the full line is the total $A_{NN}$;
	  the dushed line is the $A_{NN}^{CN}$. }
\end{center}
\end{figure}

 The model \cite{zpc} takes into account the contribution of the hadron
 interaction at large  distances and leads to
 the high-energy spin-flip amplitude.
 The model gives  the large spin effects in the $hh$-elastic scattering
 and predicts  non-small effects for the $PP2PP$ experiment at RHIC
  especially in the diffraction dip domain \cite{yaf-str}.
 The additional pure CNI effects can be calculated using the Coulomb-nuclear
  phase \cite{prd-sum}. These polarization effects will be present
  at RHIC energy, even though
  $F_{h}^{+-} \rightarrow 0$
  at high energy. Our model calculations show on Fig.3 for both
  cases.

  The model gives the standard $t$-dependence of $Re F_{h}{++}$ and
  $Im F_{h}{++}$.
  Instead of it, in a convenient parameterization of both the modulus
  and the phase
 one can  obtain the alternative case, in which $Im F_{h}(s,t)$
 has the zero at small $t$ (for details, see
\cite{kun3}). Such an approach enables one to specify
the elastic hadron scattering amplitude $F_{h}(s,t)$
directly from the elastic scattering data.
The difference between the phases leads either to
central or peripheral distributions of elastic hadron scattering
in the impact parameter space.
The obtained form of $A_{N}^{CN}$ at small momentum transfers
differs for the two variants beginning at $|t| > 0.05$ GeV$^2$ (Fig.4 a).
The difference reaches $2 \%$ at $-t = 0.15 \ GeV^2$ and,
in principle, can be measured in an accurate experiment.
Now let us calculate the Coulomb-hadron interference effect
- $A_{N}^{CN}$
in the two alternatives for higher $|t|$:
   (i) the diffraction dip is created by the "zero" of the $Im F_{h}(s,t)$
       part of the scattering amplitude and $Re F_{h}(s,t)$ fills it;
   (ii) the diffraction dip is created by the "zero" of the $Re F_{h}(s,t)$
       part of the scattering amplitude and  $Im F_{h}(s,t)$ fills it.
  The results are shown in Fig. 4 (b) for $\sqrt{s} = 50$ GeV and
  in Fig. 4 (c) at $\sqrt{s}= 500$  GeV.

\begin{figure}
\vspace{-8mm}
\begin{center}
\begin{tabular}{ccc}
\mbox{\epsfig{figure=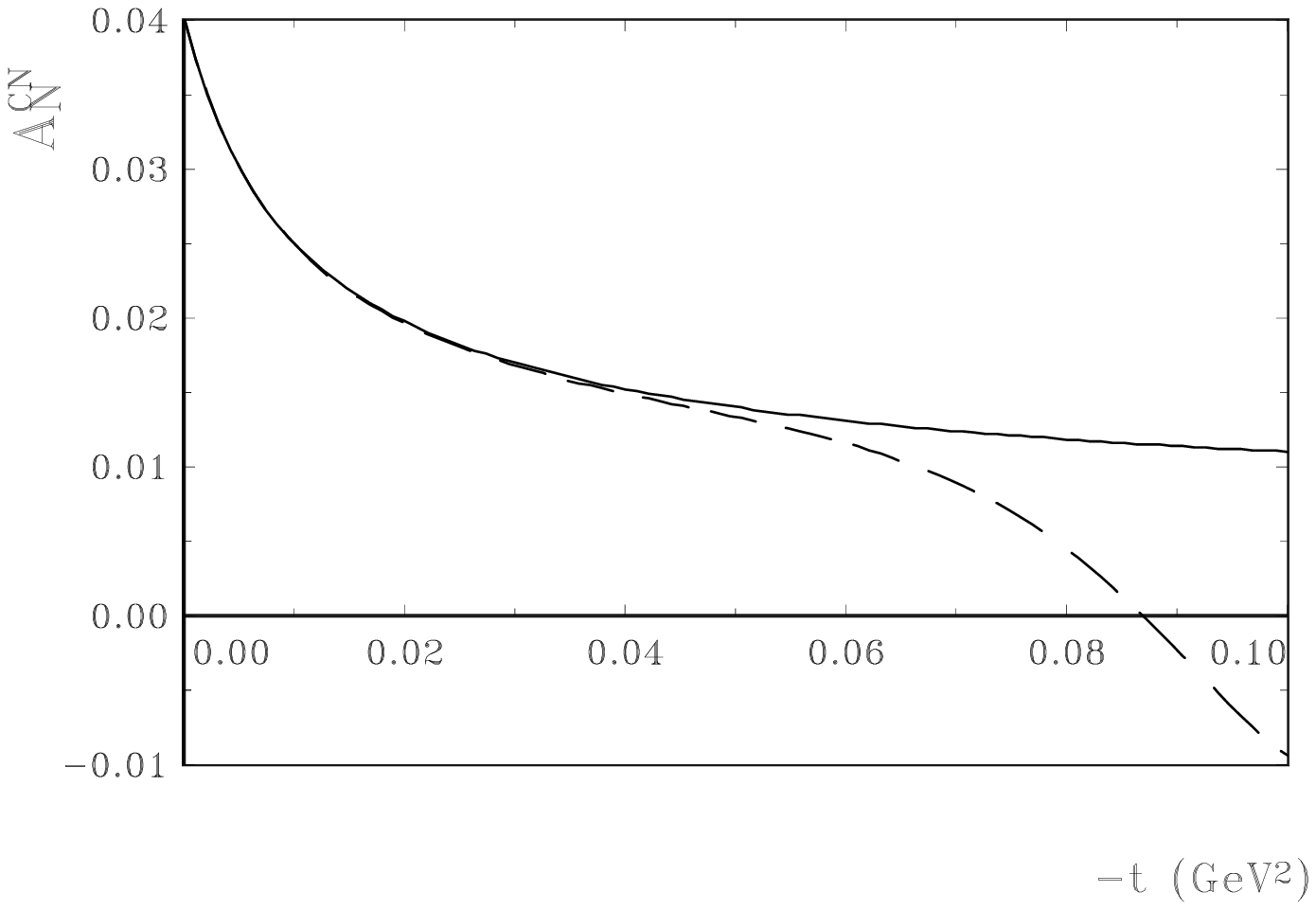,width=4.5cm,height=4cm}}&
\mbox{\epsfig{figure=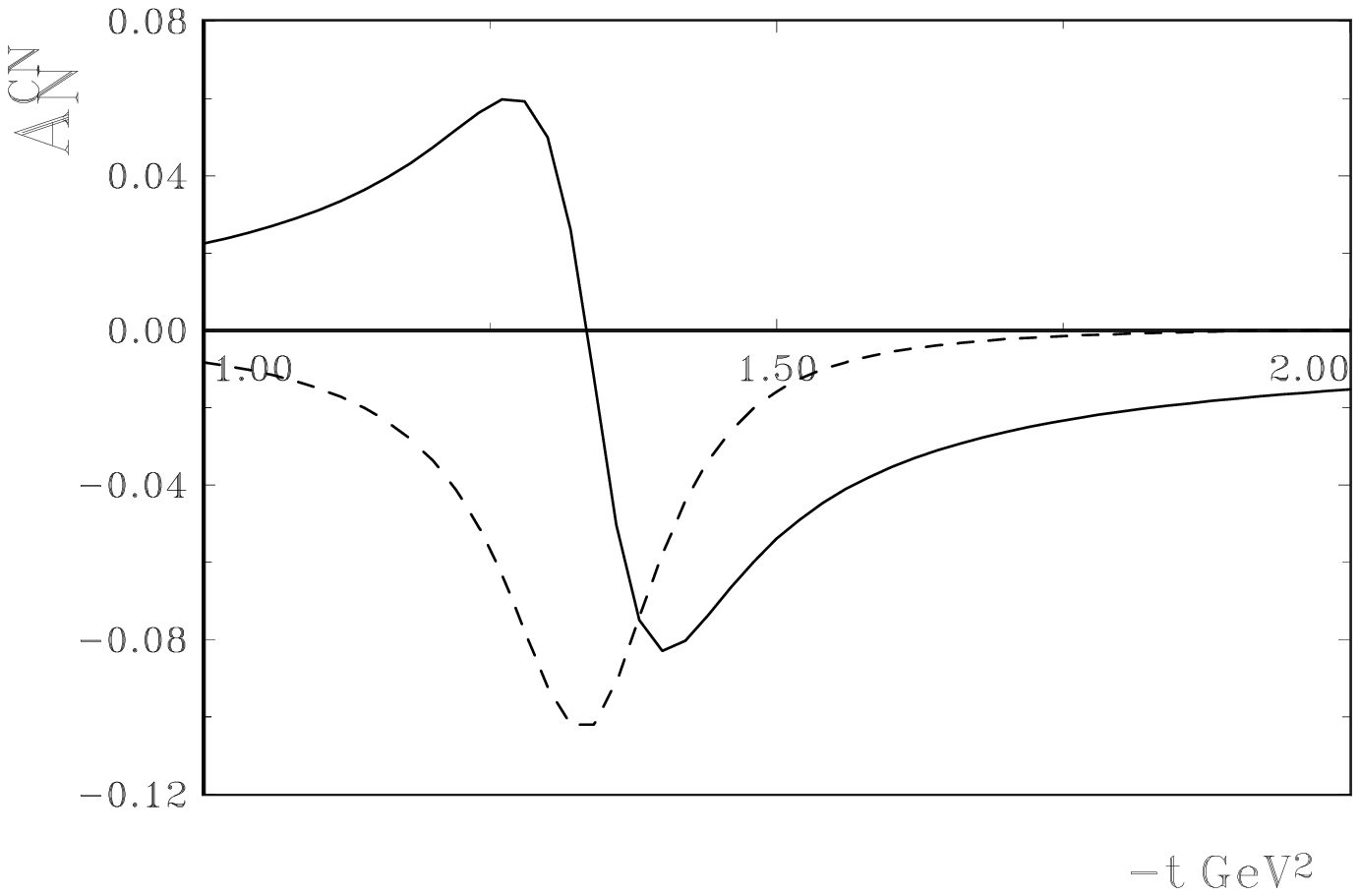,width=4.5cm,height=4cm}}&
\mbox{\epsfig{figure=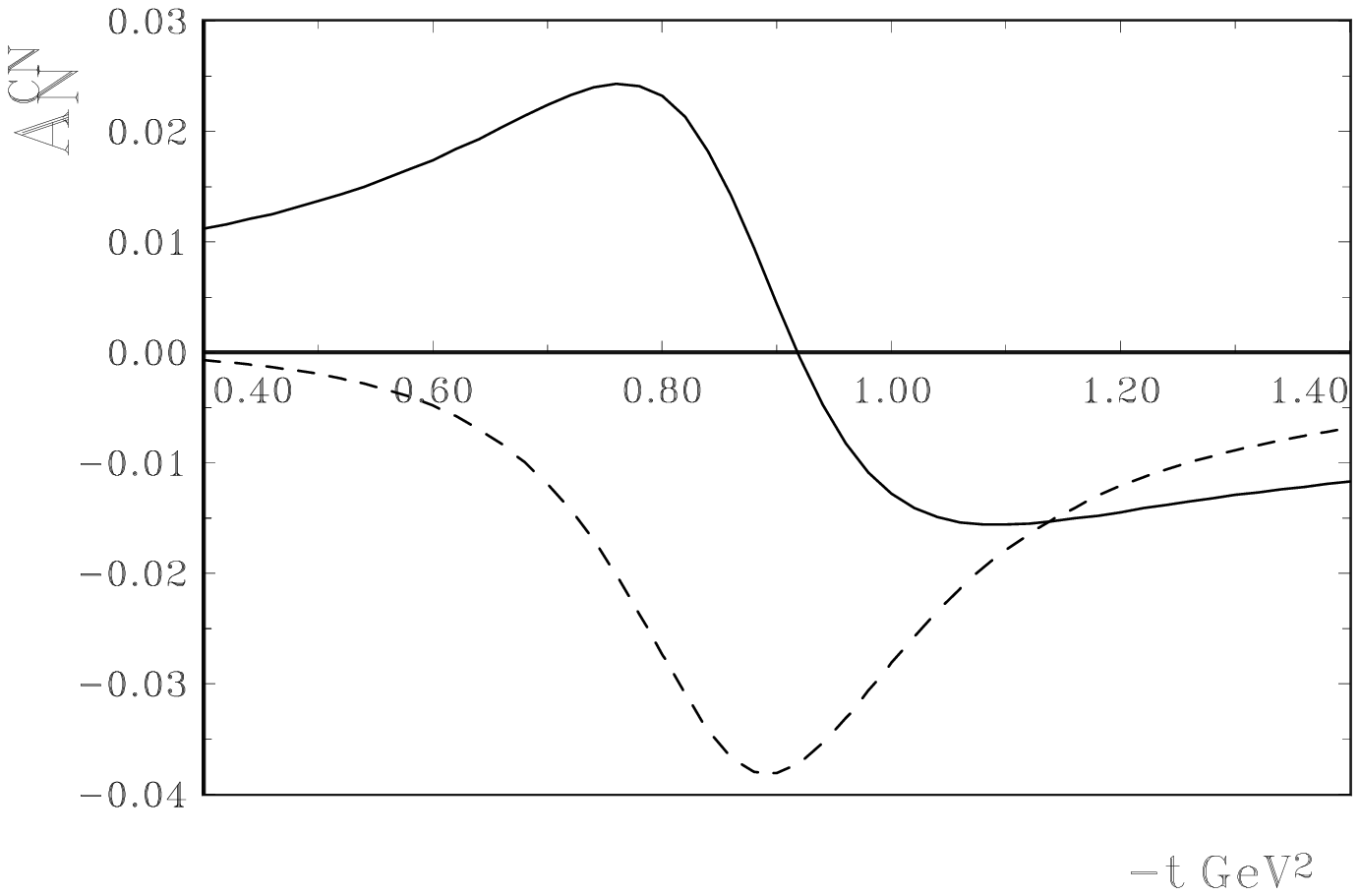,width=4.5cm,height=4cm}} \\
{\bf (a)} & {\bf (b)} &  {\bf (c)}
\end{tabular}
\caption{
({\bf a})
  $A_{N}^{CN}$  at $\sqrt{s}=50 \ GeV$
  and small $t$ for two models.
({\bf b}) and ({\bf c})
   $A_{N}^{CN}$
  at $\sqrt{s}=50 $ GeV and  $\sqrt{s}=500 \ GeV $ in the region of the dip
    (the solid line corresponds to the model I with zero of $\Im F_{h}$
   at dip; the dashed line shows the variant II, with the
   zero of the $Re F_{h}$ at the dip).}
\end{center}
\end{figure}

\end{document}